\documentclass[twocolumn,showpacs,preprintnumbers,nofootinbib,amsmath,amssymb]{revtex4}

\usepackage{euscript,amssymb,amsmath}

\usepackage{graphicx}
\usepackage{epsfig}

\newcommand{\be}[1]{\begin{equation}\label{#1}}
\newcommand{\ee}{\end{equation}}
\newcommand{\ba}[1]{\begin{eqnarray}\label{#1}}
\newcommand{\ea}{\end{eqnarray}}
\newcommand{\rf}[1]{(\ref{#1})}
\newcommand{\nn}{\nonumber}

\begin{document}

\title{Multidimensional gravity in non-relativistic limit}

\author{Maxim Eingorn}\email{maxim.eingorn@gmail.com}  \author{Alexander Zhuk}\email{ai_zhuk2@rambler.ru}
\affiliation{Astronomical Observatory and Department of
Theoretical Physics, Odessa National University, Street Dvoryanskaya 2, Odessa 65082, Ukraine}

%
%
%

%
\begin{abstract}
It is found the exact solution of the Poisson equation for the multidimensional space with topology  $M_{3+d}=\mathbb{R}^3\times T^d$.
This solution describes smooth transition from the newtonian behavior $1/r_3$ for distances bigger than periods of tori (the extra dimension sizes) to multidimensional behavior $1/r^{1+d}_{3+d}$ in opposite limit.  In the case of one extra dimension $d=1$, the gravitational potential is expressed via compact and elegant formula. It is shown that the corrections to the gravitational constant in the Cavendish-type experiment can be within the measurement accuracy of Newton's gravitational constant $G_N$. It is proposed models where the test masses are smeared over some (or all) extra dimensions. In 10-dimensional spacetime with 3 smeared extra dimensions, it is shown that the  size of 3 rest extra dimensions can be enlarged up to submillimeter for the case of 1TeV fundamental Planck scale $M_{Pl(10)}$. In the models where all extra dimensions are smeared, the gravitational potential exactly coincides with the newtonian one.  Nevertheless, the hierarchy problem can be solved in these models.
\end{abstract}

\maketitle

\vspace{.5cm}



{\em Introduction}\quad There are two well-known problems which are related to each other. They are the discrepancies in gravitational
constant experimental data and the hierarchy problem. Discrepancies (see e.g. Figure 2 in the "CODATA Recommended Values of the Fundamental
Constants: 2006") are usually explained by extreme weakness of gravity. It is very difficult to measure the Newton's gravitational constant $G_N$.
Certainly, for this reason geometry of an experimental setup can effect on data. However, it may well be that, the discrepancies can also be
explained (at least partly) by underlying fundamental theory.  Formulas for an effective gravitational constant following from such
theory can be sensitive to the geometry of experiments.  For example, if correction to the Newton's gravitational potential has the form
of Yukawa potential, then the force due to this potential at a given minimum separation per unit test-body mass is least for two spheres and
greatest for two planes (see e.g.\cite{ISL}). The hierarchy problem - the huge gap between the electroweak scale $M_{EW}\sim 10^3$GeV and the Planck scale $M_{Pl(4)} =1.2\times 10^{19}$GeV - can be also reformulated in the following manner: why is gravity
so weak? The smallness of $G_N$ is the result of relation $G_N = M_{Pl(4)}^{-2}$. The natural explanation was proposed in \cite{large,ADD}:
the gravity is strong: $G_{\mathcal{D}} = M_{Pl(\mathcal{D})}^{-(2+d)}\sim M_{EW}^{-(2+d)}$ and it happens in ($\mathcal{D}=4+d$)-dimensional spacetime.
It becomes weak when gravity is "smeared" over large extra dimensions: $G_N \sim G_{\mathcal{D}}/V_d$ where $V_d$ is a volume of internal space.
To shed light on both of these problems from new standpoint we intend to investigate multidimensional gravity in non-relativistic limit.


{\em Multidimensional gravitational potentials}\quad It is of interest to generalize the well-known Newton's gravitational
potential $\varphi(r_3) = -G_N m /r_3$ ($r_3 = |{\bf r_3}|$ is magnitude of a
radius vector in three-dimensional space) to multidimensional case. Clearly, the result depends on topology of investigated models.
We consider models where $(D=3+d)$-dimensional spatial part of factorizable geometry is defined on a product manifold
$M_D=\mathbb{R}^3\times T^{d}$.
$\mathbb{R}^3$ describes three-dimensional flat external (our) space and $T^{d}$ is a torus
which corresponds to $d$-dimensional internal space with volume $V_d$. Let  $b\sim V^{1/d}_{d}$
be a characteristic size of extra dimensions.
Then, Gauss's flux theorem leads to the following asymptotes for
gravitational potential (see e.g. \cite{ADD}): $\varphi \sim 1/r_3$ for $r_3 >> b$ and $\varphi \sim 1/r^{1+d}_{3+d}$
for  $r_{3+d} << b$ where $r_{3+d}$ is magnitude of a radius vector in
$(3+d)$-dimensional space.

To get the exact expression for $D$-dimensional gravitational potential, we start with the  Poisson equation:
\be{1}
\triangle_D\varphi_D=S_DG_{\mathcal{D}}\rho_D({\bf r}_D)\, ,
\ee
where $S_D=2\pi^{D/2}/\Gamma (D/2)$ is a total solid angle (square of
$(D-1)$-dimensional sphere of a unit radius), $G_{\mathcal{D}}$ is a gravitational constant in
$(\mathcal{D}=D+1)$-dimensional spacetime
and $\rho_D({\bf r}_D)=m\delta(x_1)\delta(x_2)...\delta(x_D)$. In the case of topology
$\mathbb{R}^D$, Eq. \rf{1} has the following solution:
\be{2}
\varphi_D({\bf r}_D)=-\frac{G_{\mathcal{D}}m}{(D-2)r_D^{D-2}}\, ,\quad D\geq 3.
\ee
This is the unique solution of Eq. \rf{1} which satisfies the boundary condition:
$\lim\limits_{r_D\rightarrow+\infty}\varphi_D({\bf r}_D)=0$. Gravitational constant
$G_{\mathcal{D}}$ in \rf{1} is normalized in such a way that the strength of gravitational
field (acceleration of a test body) takes the form: $-d\varphi_D / d r_D = - G_{\mathcal{D}}m/r^{D-1}_D$.

If topology of space is  $\mathbb{R}^3\times T^{d}$, then it is natural to impose periodic boundary
conditions in the directions of the extra dimensions:
$\varphi_D({\bf r}_3,\xi_1,\xi_2,\ldots, \xi_i,\ldots ,\xi_{d})=
\varphi_D({\bf r}_3,\xi_1,\xi_2,\ldots, \xi_i +a_i,\ldots ,\xi_{d}), \quad i=1,\ldots ,d$, where
$a_i$ denotes a period in the direction of the extra dimension $\xi_i$. Then, Poisson equation has solution
(cf. also with \cite{ADD,CB}):
\ba{3}
&{}&\varphi_D({\bf r}_3,\xi_1,...,\xi_{d})=-\frac{G_N m}{r_3}\nn \\
&\times&\sum\limits_{k_1=-\infty}^{+\infty}...\sum\limits_{k_{d}=-\infty}^{+\infty}
\exp\left[-2\pi\left(\sum\limits_{i=1}^{d}\left(\frac{k_i}{a_i}\right)^2\right)^{1/2}r_3\right]\nn \\
&\times&\cos\left(\frac{2\pi
k_1}{a_1}\xi_1\right)...\cos\left(\frac{2\pi k_{d}}{a_{d}}\xi_{d}\right)\, .
\ea
To get this result we, first, use the formula
$\delta(\xi_i)=\frac{1}{a_i}\sum_{k=-\infty}^{+\infty}\cos\left(\frac{2\pi k}{a_i}\xi_i\right)$ and, second,
put the following relation between gravitational constants in four- and $\mathcal{D}$-dimensional spacetimes:
\be{4}
\frac{S_D}{S_3}\cdot\frac{G_{\mathcal{D}}}{\prod_{i=1}^{d}a_i}=G_N\, .
\ee
The letter relation provides correct limit when all $a_i \to 0$. In this limit zero modes $k_i=0$ give the main
contribution and we obtain $\varphi_D({\bf r}_3,\xi_1,...,\xi_{d})\rightarrow-G_N m/r_3$. Eq. \rf{4} was
widely used in the concept of large extra dimensions which gives possibility to solve the hierarchy problem \cite{ADD,large}.
It is also convenient to rewrite \rf{4} via fundamental Planck scales:
\be{5}
\frac{S_D}{S_3}\cdot M_{Pl(4)}^{2} = M_{Pl(\mathcal{D})}^{2+d}\prod_{i=1}^{d}a_i\, ,
\ee
where $M_{Pl(4)}= G_N^{-1/2} =1.2\times 10^{19}$GeV and $ M_{Pl(\mathcal{D})}\equiv G_{\mathcal{D}}^{-1/(2+d)}$ are
fundamental Planck scales in four and $\mathcal{D}$ spacetime dimensions, respectively.

In opposite limit when all $a_i \to +\infty$ the sums in Eq. \rf{3} can be replaced by integrals. Using the standard integrals
(e.g. from \cite{PBM}) and relation \rf{4}, we can easily show  that, for example, in particular cases $d=1,2$ we get
desire result: $\varphi_D({\bf r}_3,\xi_1,\ldots ,\xi_d)\rightarrow-G_{\mathcal{D}} m/[(D-2)\; r_{3+d}^{1+d}]$.

{\em One extra dimension}\quad In the case of one extra dimension $d=1$ we can perform summation of series in Eq. \rf{3}.
To do it, we can apply the Abel-Plana formula or simply use the tables of series \cite{PBM}. As a result, we arrive at
compact and nice expression:
\be{6}
\varphi_4({\bf r}_3,\xi)=-\frac{G_N m}{r_3}\frac{\sinh\left(\frac{2\pi r_3}{a}\right)}{\cosh\left(\frac{2\pi
r_3}{a}\right)-\cos\left(\frac{2\pi\xi}{a}\right)}\, ,
\ee
where $r_3 \in [0,+\infty )$ and $\xi \in [0,a]$. It is not difficult to verify that this formula has correct asymptotes
when $r_3>>a$ and $r_4<<a$. Fig. 1 demonstrates the shape of this potential. Dimensionless variables $\eta_1\equiv r_3/a \in [0,+\infty )$
and $\eta_2\equiv \xi/a \in [0,1]$. With respect to variable $\eta_2$, this potential has two minima at $\eta_2=0,1$ and one maximum at
$\eta_2 =1/2$. We continue the graph to negative values of $\eta_2 \in [-1,1]$ to show in more detail
the form of minimum at $\eta_2=0$. The potential \rf{6} is finite for any value of $r_3$ if $\xi\neq 0,a$ and goes
to $-\infty $ as $ -1/r_4^2$ if
simultaneously $r_3 \to 0$ and $\xi \to 0,a$ (see Fig. 2). We would like to mention that in particular case $\xi=0$ formula \rf{6}
was also found in \cite{Barvinsky}.
\begin{figure}[htbp]
\includegraphics[width=2.7in,height=1.8in]{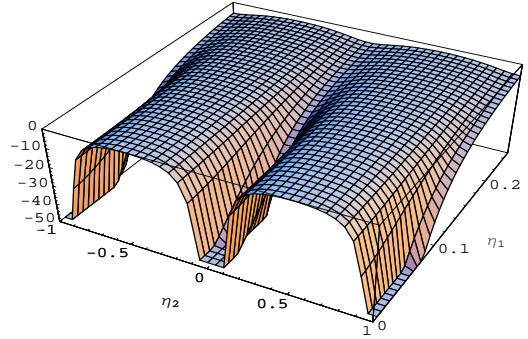}
\caption {Graph of function $\tilde \varphi (\eta_1,\eta_2) \equiv \varphi_4({\bf r}_3,\xi)/(G_N m/a)=
-\sinh (2\pi \eta_1)/[\eta_1(\cosh(2\pi\eta_1)-\cos(2\pi\eta_2))]$. \label{potential}}
\end{figure}

\begin{figure}[htbp]
\includegraphics[width=2.5in,height=1.4in]{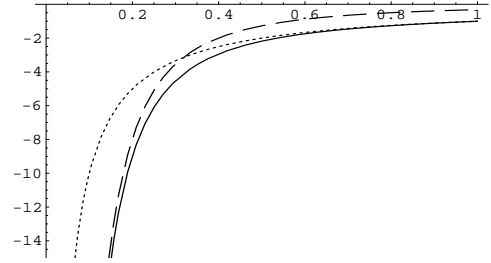}
\caption {Section $\xi =0$ of potential \rf{6}. Solid line is $\tilde \varphi (\eta_1,0) =
-\sinh (2\pi \eta_1)/[\eta_1(\cosh(2\pi\eta_1)-1)]$ which goes to $-1/\eta_1$ (dotted line) for $\eta_1 \rightarrow +\infty$
and to $-1/(\pi \eta_1^2)$ (dashed line) for $\eta_1 \rightarrow 0$. \label{asympt}}
\end{figure}

Having at hand formulas \rf{3} and \rf{6}, we can apply it for calculation of some elementary physical problems and compare
obtained results with known newtonian expressions.
Some of these calculations can be found in our preprint \cite{EZ}.
For a working approximation, it is usually sufficient to summarize in \rf{3} up to the first Kaluza-Klein modes $|k_i|=1\, (i=1,\ldots ,d)$. Then, the terms with the biggest periods $a_i$ give the main contributions.  If all test bodies
are on the same brane ($\xi_i=0$) we obtain:
\ba{7a}
&{}&\varphi_D({\bf r}_3,\xi_1=0,...,\xi_{d}=0) \equiv \varphi_D (r_3)\approx -\frac{G_N m}{r_3}\nn \\
&\times& \left[1+\alpha \exp\left(-\frac{r_3}{\lambda}\right)\right]\, ,
\ea
where $\alpha =2s \; (1 \le s\le d)\, $, $\lambda =a/(2\pi)$ and $s$ is a number of extra dimensions with periods of tori $a_i$ which
are equal (or approximately equal) to $a=\max a_i$. If $a_1=a_2= \ldots = a_d=a$, then $s=d$. Thus, the correction to Newton's
potential has the form of Yukawa potential. It is now customary to interpret tests of gravitational inverse-square law (ISL) as setting limits on additional Yukawa contribution. The overall diagram of the experimental constraints can be found in \cite{ISL}  (see Figure 5) and we shall
use these data for limitation $a$ for given $\alpha$.

In this approximation, the gravitational force between two spheres with masses $m_1,m_2$, radiuses $R_1,R_2$ and distance $r_3$ between the centers
of the spheres reads:
\be{8b}
F = -\frac{G_{N(eff)} m_1m_2}{r_3^2}\, ,
\ee
where
\ba{8a}
&\phantom{}&G_{N(eff)}(r_3)\approx G_N \left\{1+\frac{9}{2}s\left(\frac{a}{2\pi R_1}\right)^2\left(\frac{a}{2\pi R_2}\right)^2\right.\nn \\
&\times& \left.\frac{2\pi
r_3}{a}\exp\left[-\frac{2\pi}{a}(r_3-R_1-R_2)\right]\right. \equiv G_N(1+\delta_G).
\ea
Here, we used conditions: $r_3\ge R_1+R_2$ and $R_1,R_2\gg a/2\pi$.

{\em Smeared extra dimensions}\quad In what follows, we consider asymmetrical extra dimension model (cf. \cite{asymm}) with topology
\be{9a}
M_D=\mathbb{R}^3\times T^{d-p}\times T^p \, , \quad p\le d\, ,
\ee
where we suppose that $(d-p)\, $ tori have the same "large" period $a$ and $p$ tori have "small" equal periods $b$. In this case, the fundamental
Planck scale relation \rf{5} reads
\be{10a}
\frac{S_D}{S_3}\cdot M_{Pl(4)}^{2} = M_{Pl(\mathcal{D})}^{2+d}a^{d-p}\, b^p\, .
\ee
Additionally, we assume that test bodies are uniformly smeared/spreaded over small extra dimensions. Thus, test bodies have a finite thickness in small extra dimensions (thick brane approximation). For short, we shall call such small extra dimensions as "smeared" extra dimensions. If $p=d$ then all
extra dimensions are smeared.

It is not difficult to show that the gravitational potential does not feel smeared extra dimensions. We can prove this statement by three different methods. First, we can directly solve D-dimensional Poisson equation \rf{1} with
the periodic boundary conditions for the extra dimensions $\xi_{p+1},\ldots ,\xi_d$ and the mass density $\rho = \left(m/\prod_{i=1}^p a_i\right)
\delta({\bf r}_3)\delta(\xi_{p+1})...\delta(\xi_d)$. Second, we can average solutions \rf{3} and \rf{6} over dimensions $\xi_1, \ldots ,\xi_p$ and
take into account that $\int_0^a \cos(2\pi k\xi/a)d\xi =0$. In particular case of one extra dimensional, we can also show that
\ba{11a}
&-&\frac{G_N m}{ar_3}\sinh\left(\frac{2\pi r_3}{a}\right)\int\limits_0^a\left[\cosh\left(\frac{2\pi
r_3}{a}\right)\right.\nn \\
&-&\left.\cos\left(\frac{2\pi\xi}{a}\right)\right]^{-1}d\xi = -\frac{G_N m}{r_3} \, .
\ea
Finally, it is clear that in the case of test masses smeared over extra dimensions, the gravitational field vector ${\bf E}_D= -\nabla_D \varphi_D$
does not have components with respect to extra dimensions: ${\bf E}_D=E_D {\bf n}_{r_3}$. Thus, applying the Gauss's flux theorem to the Poisson
equation, we obtain: $E_D(r_3)=-G_Nm/r_3^2\; \rightarrow \; \varphi_D(r_3)=-G_Nm/r_3$. Therefore, all these 3 approaches show that in the case of $p\, $ smeared extra dimensions the wave numbers $k_1,\ldots ,k_p$ disappear from Eq. \rf{3} and we should perform summation only with respect to $k_{p+1}, \ldots ,k_d$.

{\em Effective gravitational constant}\quad Coming back to the effective gravitational constant \rf{8a} in the case of topology \rf{9a} with $p\, $ smeared extra dimensions, $s$ in Eq. \rf{8a} is replaced by $(d-p)$.
Now, we want to evaluate the corrections $\delta_G$ to the Newton's gravitational constant and to estimate their possible influence on the experimental data.
As it follows from Figure 2 in the CODATA 2006, the most precise values of $G_N$ were obtained in the University
Washington and the University Z\"urich experiments \cite{U-wash,U-zur}. They are  $G_N/10^{-11}{\mbox m}^3{\mbox kg}^{-1}{\mbox s}^{-2} =
6.674215\pm 0.000092$,
and $6.674252\pm 0.000124$, respectively. Let us consider two particular examples: $(\mathcal{D}=5)$-dimensional model  with $d=1, p=0\; \rightarrow \alpha = 2$ and
$(\mathcal{D}=10)$-dimensional model  with $d=6, p=3\; \rightarrow \alpha = 6$. For these
values of $\alpha$, Figure 5 in \cite{ISL} gives the upper limits for $\lambda=a/(2\pi)$ correspondingly $\lambda  \approx 2\times 10^{-2}$cm and
$\lambda  \approx 1.3\times 10^{-2}$cm. To calculate $\delta_G$, we take parameters of the Moscow experiment \cite{Moscow}: $R_1\approx 0.087$cm for a platinum ball with the mass $m_1=59.25\times 10^{-3}$g, $R_2\approx 0.206$cm for a tungsten ball with the mass $m_2=706\times 10^{-3}$g and $r_3=0.3773$cm. For both of these models we obtain $\delta_G \approx 0.0006247$ and $\delta_G \approx 0.0000532$, respectively. Both of these values are very close to the measurement accuracy of $G_N$ in \cite{U-wash,U-zur}. So, if the same accuracy can be achieved in the Moscow-type experiments, then, changing values of $R_{1,2}$ and $r_3$, we can reveal extra dimensions or obtain experimental limitations on considered models.

{\em Model: $\mathcal{D}=10$ with $d=6, p=3$}\quad Let us consider in more detail $(\mathcal{D}=10)$-dimensional model  with 3 smeared dimensions.
Here, we have very symmetrical with respect to a number of spacial dimensions structure: 3 our external dimensions, 3 large extra dimensions with periods $a$ and three small smeared extra dimensions with
periods $b$. For $b$ we put a limitation: $b\le b_{max} = 10^{-17}$cm which is usually taken for thick brane approximation. As we mentioned above, in the case of $\alpha =6$, for $a$ we should take a limitation $a\le a_{max} =8.2\times 10^{-2}$cm.  To solve the hierarchy problem, the multidimensional Planck scale is usually considered from 1TeV up to approximately 130 TeV (see e.g. \cite{asymm,supernova}). To make some estimates, we take
$M_{min}=1$TeV$\lesssim M_{Pl(10)}\lesssim M_{max}=50$TeV. Thus, as it follows from Eq. \rf{10a}, the allowed values of $a$ and $b$ should satisfy
inequalities:
\be{12a}
\frac{S_9}{S_3}\frac{M_{Pl(4)}^2}{M_{max}^{8}}\le a^3b^3\le \frac{S_9}{S_3}\frac{M_{Pl(4)}^2}{M_{min}^{8}}
\ee
Counting all limitations, we find allowed region for $a$ and $b$ (shadow area in Fig. \rf{trapezium}). In this trapezium, the upper horizontal and right vertical lines are the decimal logarithms of $a_{max}$ and $b_{max}$, respectively. The  right and left inclined lines correspond to
$M_{Pl(10)}=1$TeV and $M_{Pl(10)}=50$TeV, respectively. To illustrate this picture, we consider two points A and B on the line $M_{Pl(10)}=1$TeV.
Here, we have $a=0.82\times 10^{-1}$cm, $b=10^{-21.5}$cm for A and $a=10^{-4}$cm, $b=10^{-18.6}$cm for B. These values of large extra dimensions $a$ are much bigger than in the standard approach $a \sim 10^{(32/6)-17}$cm $\approx 10^{-11.7}$cm \cite{large,ADD}.
\begin{figure}[htbp]
\includegraphics[width=2.5in,height=1.4in]{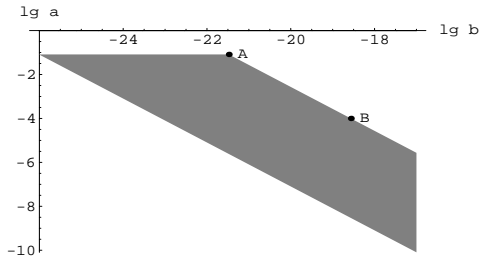}
\caption {Allowed region (shadow area) for periods of large ($a$) and smeared ($b$) dimensions in the model $\mathcal{D}=10$ with $d=6, p=3$. \label{trapezium}}
\end{figure}

{\em Model: $\mathcal{D}$-arbitrary and $d= p$}\quad In this model, the test masses are smeared over all extra dimensions.
Therefore, in non-relativistic limit, there is no deviation from the Newton's law at all. It worth of noting, that this result
does not depend on the size of smeared extra dimensions. The ISL experiments will not show any deviation from the Newton's law
without regard to the size $b$ (see also Eq. \rf{8a} where $s=d-p=0$). Similar reasoning are also applicable to Coulomb's law.  It is necessary to suggest other experiments
which can reveal the multidimensionality  of our spacetime. Nevertheless, we can solve the hierarchy problem in this model
because Eq. \rf{10a} (where $d=p$) still works.  For example, in the case of bosonic string dimension $\mathcal{D}=26$
we find $M_{Pl(26)}\approx 31$TeV for $b =10^{-17}$cm. In the case $\mathcal{D}=10$ we get $M_{Pl(10)}\approx 30$TeV for $b =5.59\times 10^{-14}$cm.

{\em Conclusions}\quad We have considered generalization of the Newton's potential to the case of extra dimensions where multidimensional space
has topology $M_D=\mathbb{R}^3\times T^{d}$. We obtained the exact solution which describes smooth transition from the newtonian behavior $1/r_3$ for distances bigger than periods of tori (the extra dimension sizes) to multidimensional behavior $1/r^{D-2}_D$ in opposite limit.  In the case of one extra dimension,
the gravitational potential is expressed via compact and elegant formula \rf{6}.

As an Yukawa potential approximation, it was shown that the corrections to the gravitational constant in the Cavendish-type experiment can be within the measurement accuracy of $G_N$. It may reveal the extra dimensions or provide experimental limitations on parameters of multidimensional models.

Then, we proposed models where test masses can be smeared over extra dimensions. In this case, the gravitational potential does not feel smeared dimensions. The number of smeared dimensions can be equal or less than the total number of the extra dimensions.
Such approach opens new remarkable possibilities.

For example in the case $\mathcal{D}=10$ with
3 large and 3 smeared extra dimensions and $M_{Pl(10)} =1$TeV, the large extra dimensions can be as big as the upper limit established by the ISL experiments for $\alpha =6$, i.e. $a\approx 0.82\times 10^{-1}$cm. This value of $a$ is in many orders of magnitude bigger than the rough estimate $a \approx 10^{-11.7}$cm obtained from the fundamental Planck scale relation of the form of Eq. \rf{5}.

The limiting case where all extra dimensions are smeared is another interesting example. Here, there is no deviation from the Newton's law at all.
Nevertheless, the hierarchy problem can be solved in this model.


\indent \indent A. Zh. acknowledges the hospitality
of the Theory Division of CERN and the High Energy, Cosmology and Astroparticle Physics Section of the ICTP
during preparation of this work.
This work was supported in part by the
"Cosmomicrophysics" programme of the Physics and Astronomy
Division of the National Academy of Sciences of Ukraine.




\begin{thebibliography}{}

\bibitem{ISL}
E.G. Adelberger, B.R. Heckel and A.E. Nelson, Ann. Rev. Nucl. Part. Sci. {\bf 53}, 77 (2003); arXiv:hep-ph/0307284.
\bibitem{large}
N. Arkani-Hamed, S. Dimopoulos and G. Dvali, Phys. Lett. B {\bf 429}, 263 (1998); hep-ph/9803315.
I. Antoniadis, N. Arkani-Hamed, S. Dimopoulos and G. Dvali, Phys. Lett. B {\bf 436}, 257 (1998); arXiv:hep-ph/9804398.
\bibitem{ADD}
 N. Arkani-Hamed, S. Dimopoulos and G. Dvali, Phys. Rev. D {\bf 59}, 086004 (1999);
arXiv:hep-ph/9807344.
\bibitem{CB}
 P. Callin and C.P. Burgess,  Nucl. Phys. B {\bf 752}, 60 (2006);
arXiv:hep-ph/0511216.
\bibitem{PBM}
A.P. Prudnikov, Yu.A. Brychkov and O.I. Marichev, {\it Integrals and Series, vol. 1: Elementary Functions},
(Gordon and Breach Science Publishers, New York, 1986).
\bibitem{Barvinsky}
A. O. Barvinsky and S. N. Solodukhin,
Nucl. Phys. B {\bf 675}, 159 (2003); arXiv:hep-th/0307011 .
\bibitem{EZ}
M. Eingorn and A. Zhuk, {\it The shape of multidimensional gravity}, 	arXiv:0905.2222.
\bibitem{asymm}
J. Lykken and S. Nandi, Phys.Lett. B {\bf 485}, 224 (2000); arXiv:hep-ph/9908505.
\bibitem{U-wash}
J.H. Gundlach and S.M. Merkowitz, Phys.Rev.Lett. {\bf 85}, 2869 (2000); arXiv:gr-qc/0006043.
\bibitem{U-zur}
St. Schlamminger et al., Phys.Rev. D {\bf 74}, 082001 (2006); arXiv:hep-ex/0609027.
\bibitem{Moscow}
V.P. Mitrofanov and O.I. Ponomareva, Sov. Phys. JETP {bf 67}, 1963 (1988).
\bibitem{supernova}
S. Hannestad and G. Raffelt, Phys. Rev. Lett. {\bf 87}, 051301 (2001); arXiv:hep-ph/0103201.





\end{thebibliography}
\end{document}